\documentclass[preprint,12pt]{elsarticle}
\usepackage{paper}
\journal{Nuclear Instruments and Methods in Physics Research Section A}
\begin{document}
\begin{frontmatter}
\title{A 40 Gbit/s FPGA-based RoCEv2 transport for detector readout: development and performance}
\author[ihep]{Sheng Dong\corref{cor1}}
\ead{dongsheng@ihep.ac.cn}
\cortext[cor1]{Corresponding author}
\author[ihep]{Jingzhou Zhao}
\ead{zhaojz@ihep.ac.cn}
\author[ihep,ucas]{Haoxin Wang}
\ead{wanghaoxin@ihep.ac.cn}
\author[ihep]{Tingxuan Zeng}
\ead{zengtx@ihep.ac.cn}
\author[ihep]{Fei Li}
\ead{lifei@ihep.ac.cn}
\affiliation[ihep]{organization={Institute of High Energy Physics, Chinese Academy of Sciences},
  city={Beijing}, postcode={100049}, country={China}}
\affiliation[ucas]{organization={University of Chinese Academy of Sciences},
  city={Beijing}, postcode={100049}, country={China}}
\begin{abstract}
Recording the complete input of a detector trigger allows trigger algorithms to be replayed and studied offline. For the BESIII main drift chamber (MDC), this requires moving tens of Gbit/s from trigger FPGAs into server memory without loss. Remote direct memory access over Converged Ethernet (RoCEv2) lets commercial network adapters write such data into host memory. Detector readout also needs record boundaries that the receiver can check, replay from a verified copy, and host buffers that are reused only after consumption. We present a 40 Gbit/s FPGA RoCEv2 transport that provides these functions. Records are aggregated into CRC32C-checked blocks, and unacknowledged data are kept in DDR memory and verified before replay. Acknowledgment-aware retention writes to DDR only the data that are not acknowledged in time. The transport runs on a ZCU102 board and on the 9U VME Trigger Core Board (TCB). With records of 128 B to 4 KiB it comes within 0.4\% of the packetization limit of the link, delivering 36.9 Gbit/s of payload at a path MTU of 1024 B, and its block latency is set by the time needed to fill a block. On a direct link, acknowledgment-aware retention cuts DDR writes for large blocks from 106\% to 0.03\% of the payload, and injected DDR access errors stay within their queue pair. On the TCB, three 16.4 TB transfers through two switches lost at most five packets each, all recovered by replay. An installation at the MDC is under test.
\end{abstract}

\end{frontmatter}
\FloatBarrier
\section{Introduction}
\label{sec:1}

The BESIII main drift chamber (MDC) provides charged-particle tracking information and input to the first-level trigger system~\cite{Ablikim2010BESIII}. Access to its trigger inputs supports offline replay, trigger-algorithm studies and investigation of anomalous detector or electronics behavior, and the MDC trigger-electronics upgrade includes an additional readout path and data formats for this purpose~\cite{Wang2026MDCReadout}. The path needs a complete transport chain from FPGA electronics to server memory, in which the physical link, remote direct memory access (RDMA) protocol processing, record handling, buffering and host reception work together. Transport acknowledgments, completed writes to retention memory and the end of host processing occur at different points in this chain, and the design has to tie each to its own decision: releasing retained data, replaying them, or reusing a host buffer.

The Trigger Core Board (TCB) is a 9U VME (Versa Module Eurocard) development platform based on XCKU060 or XCKU115 FPGAs, designed for detector subtriggers and the global trigger. Its readout interface sends MDC trigger inputs to data acquisition (DAQ) servers while the trigger and control functions continue to run~\cite{IHEPTDAQDRD72026}.

RDMA allows a network interface to write into preregistered host memory. RDMA over Converged Ethernet version 2 (RoCEv2) supports communication between FPGAs and commercial RDMA network interface cards (RNICs) over Ethernet~\cite{IBTA2014RoCEv2}. Prior work demonstrates FPGA-to-RNIC transmission for FELIX~\cite{Vasile2022FELIXRDMA}, develops continuous streaming and multiple-receiver operation~\cite{Vasile2025RDMAStreaming}, and combines FPGA processing with RDMA acquisition for two-dimensional X-ray detectors~\cite{Mansour2021RASHPA}. Earlier stages of the FPGA RDMA development to which this work belongs appeared in a journal article~\cite{Xu2026BEERDMA} and in a conference contribution on the 40 Gbit/s RoCEv2 protocol stack with summary application results~\cite{Dong2026CEPCRoCE}. The journal article describes a separate 100 Gbit/s prototype for the back-end electronics of the Circular Electron Positron Collider (CEPC) on a VCU118 board, built on a custom single-queue-pair (QP) RDMA SEND engine without retransmission. The 40 Gbit/s transport reported here uses a different protocol core and transfer mode.

RoCEv2 leaves three interfaces undefined for detector readout, and the design specifies each:
\begin{enumerate}
\item Records, blocks and notifications. The FPGA forms blocks carrying valid lengths, record counts, sequence state and cyclic redundancy check (CRC32C) values. The receiver combines this metadata with completion information to locate and check every received block.
\item Replay of retained data. Acknowledgment (ACK)-aware writeback, completed-write barriers and metadata/payload checks govern release from double-data-rate (DDR) memory to the replay path.
\item Host memory and session lifetime. The host software manages the RDMA resources and the FPGA session, and returns receive credits only when the selected consumer has finished with a buffer.
\end{enumerate}

The RoCEv2 protocol core is taken from the open-source transmit-only endpoint 100G-verilog-RoCEv2-lite~\cite{Bortolato2025RoCEv2Lite}, whose Ethernet, IP and UDP layers derive from verilog-ethernet~\cite{Forencich2020VerilogEthernet}. The imported core already provides RDMA WRITE and SEND with Immediate, retransmission with on-chip buffering, a UDP connection manager, a multi-QP state table with round-robin arbitration, and insertion of the invariant CRC (ICRC). This work adds the 40GBASE-R port, the record aggregator with its block format and notification contract, DDR retention memory with acknowledgment-aware writeback and validated replay, CRC32C integrity checks, class-driven multi-QP operation, session management with orderly closure, priority-based flow control (PFC) reception, an experimental rate controller based on Data Center Quantized Congestion Notification (DCQCN), and the host receiver. The contribution is the integration and validation of these functions for detector-record readout. The design is evaluated on a ZCU102 evaluation board, which also carries the multi-QP, flow-control and congestion-control extensions, and as a single-QP design on two TCBs.

\FloatBarrier
\section{Requirements and architecture}
\label{sec:2}

Sending each short detector record as its own message generates frequent packets and completion notifications, while concatenating bytes into larger messages does not by itself show that the data are complete. Data are instead organized at four levels: records, transport blocks, RDMA messages and network packets. The block size sets the transfer granularity, and record order is preserved at every level. The link rate of 40 Gbit/s matches the ports of the TCB and leaves margin over the reported per-board MDC rate of 13.4 Gbit/s~\cite{IHEPTDAQDRD72026}. Table~\ref{tab:1} lists the readout constraints, the design response to each and the parameters that the MDC integration must supply.

\begin{table}[htbp]
\small
\caption{Readout constraints, the design response to each, and the parameter that the MDC integration must supply.}
\label{tab:1}
\begin{tabularx}{\linewidth}{@{}YYY@{}}
\toprule
Constraint & Design response & MDC parameter \\
\midrule
Packet and notification rate of short records & Aggregate records into checked blocks & Record sizes and arrival pattern \\
Recoverable order and record boundaries & Keep byte order; sequence, count and length in the header & Self-describing variable-length format \\
Missing feedback and storage errors & Retain unacknowledged data; check before replay & Expected fault and stall range \\
Finite buffering and waiting time & Backpressure, tail flush, selectable block target & Source pausability, buffers, latency budget \\
Integration into trigger electronics & Separate the subsystem from board Ethernet and DDR interfaces & Integrated firmware and input coverage \\
\bottomrule
\end{tabularx}
\end{table}

\begin{figure}[htbp]
\centering
\includegraphics[width=\linewidth]{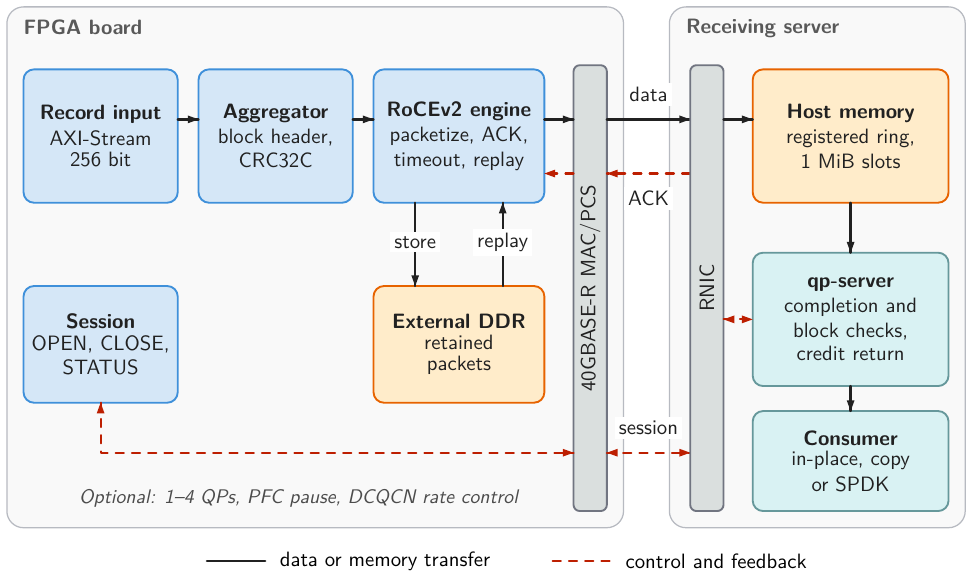}
\caption{Logical architecture of the 40 Gbit/s FPGA-based RoCEv2 transport. Colors distinguish FPGA logic (blue), memory (orange), host software (teal) and network interfaces (grey); dashed red lines denote control and feedback. The optional multi-QP, PFC and DCQCN extensions are built only on ZCU102; the TCB design uses a single QP.}
\label{fig:1}
\end{figure}

One complete AXI-Stream frame represents one source record. The application interface has twice the capacity of the link. The aggregator organizes records into transport blocks, each written into a host ring buffer by one RDMA Write with Immediate. The sender segments the message according to the path maximum transmission unit (PMTU). The receiver uses completion notifications to locate slots and checks the block header, sequence number and payload. Figure~\ref{fig:1} shows the data path, retransmission storage and session management.

The application interface propagates backpressure through TREADY. A source that cannot pause still requires application-side buffering and an overflow policy. Aggregation also introduces waiting time: with constant input rate $r_{\mathrm{in}}$, the first records of a block wait on a time scale of $L/r_{\mathrm{in}}$ for a target payload of $L$ bytes. The evaluation uses a target of 1016 KiB, which, with the block header, fits a 1 MiB host slot. At the reported per-board MDC rate of 13.4 Gbit/s, filling such a block takes about 0.62 ms. A finite batch flushes its partial tail block with the final record. For sparse detector input, the waiting time must be bounded by a flush rule, such as a timeout, which the present interface lacks.

\FloatBarrier
\section{Development and implementation}
\label{sec:3}

\subsection{Board integration and host software}

The protocol subsystem is separated from the board implementation at the 256-bit Ethernet AXI-Stream and DDR memory interfaces. The board wrapper supplies the media access control/physical coding sublayer (MAC/PCS), transceivers, clocks and resets, and an AMD Ethernet IP core provides the physical link~\cite{AMD2025PG211}. Four SFP+ lanes on ZCU102 form one 40GBASE-R port.

The receiving application sets up the RDMA resources and starts polling the completion queue (CQ) before it asks the FPGA to open the session. For RDMA Write with Immediate, a posted receive work request supplies only a receive credit for the completion notification. The receiver decodes the Immediate slot and generation to locate the data written into the ring, and validates block format, valid length, CRC32C and per-QP notification order. Header or payload CRC failures, invalid notifications and routing errors stop the session. Rejected blocks are not counted as accepted data or passed to consumers.

When receive credits are returned depends on the consumer. In-place reception returns credit after the receive checks, and a storage prototype based on the Storage Performance Development Kit (SPDK) returns it only after the NVMe writes have completed~\cite{SPDKNVMeDriver}, so that storage delays appear to the sender as receiver-not-ready (RNR) responses rather than as overwritten buffers. During orderly shutdown, the receiver requests FPGA CLOSE while CQ processing remains active and stops polling only after the FPGA confirms that the QPs are reset.

\subsection{Record interface, aggregation and completion semantics}
\label{sec:aggregation}

Records are delimited by AXI-Stream frame boundaries and are at most 4 KiB long. The last record of a batch can force out a tail block that has not reached the target length.

\begin{figure}[htbp]
\centering
\includegraphics[width=\linewidth]{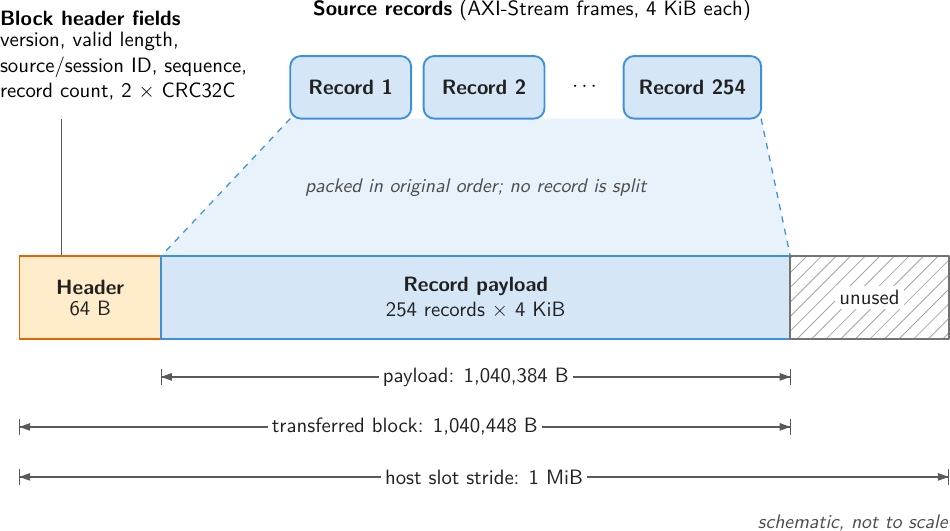}
\caption{Record aggregation into a transport block and its host-memory slot, for a full block of 254 records of 4 KiB. The block header and unused slot space are excluded from payload goodput; a final partial block is checked with its actual length and record count.}
\label{fig:2}
\end{figure}

Each block consists of a 64 B header and a payload. The header records the format version, valid length, source and session identifiers, block sequence number, record count, and header and payload CRC32C values. Figure~\ref{fig:2} shows a block of 254 records of 4 KiB in a 1 MiB host slot. The aggregator packs records consecutively in their original order and does not split a record across blocks. Because the header carries no per-record length table, the host requires either a known fixed record length or a length field within each record, and variable-length detector records must follow this rule. Each block is associated with one completion notification. The header is about 0.006\% of a full block but 1.5\% when a single 4 KiB record forms a block, so block granularity sets both header overhead and notification frequency.

\subsection{Integrity checks}

Three checks protect different parts of the data path. The ICRC covers the network packet. Separate CRC32C values cover the block header and payload for host validation. A further CRC32C covers the metadata and valid payload of each retained record in DDR. It has to be checked before replay, because an ICRC generated after a corrupted memory read would cover the corrupted bytes. The CRC32C engines process 32 bytes per clock cycle and are not in the backpressure path.

\subsection{Acknowledgment-aware retention and safe replay}
\label{sec:retention}

Data required for replay must remain available until the receiver acknowledges them. The inherited core keeps them in on-chip buffers, but at 40 Gbit/s a 5 ms retransmission timeout corresponds to about 25 MB in flight, so this design retains them in external DDR. Writing every packet to DDR would in turn consume write bandwidth comparable to the link rate. The implementation therefore first keeps transmitted data in a local buffer and relates that retained range to ACK progress. Data acknowledged before they would be written to DDR are evicted without DDR traffic. Data still needed for replay are written to DDR, with adjacent writes coalesced. In the ZCU102 and TCB builds, transmitted packets pass through a 64 KiB staging FIFO, and up to eight frames are held for at most 2048 clock cycles (6.6~$\mu$s). Frames acknowledged within that time are dropped, and the rest are written to DDR. The DDR traffic therefore depends on how quickly ACKs return.

\begin{figure}[htbp]
\centering
\includegraphics[width=\linewidth]{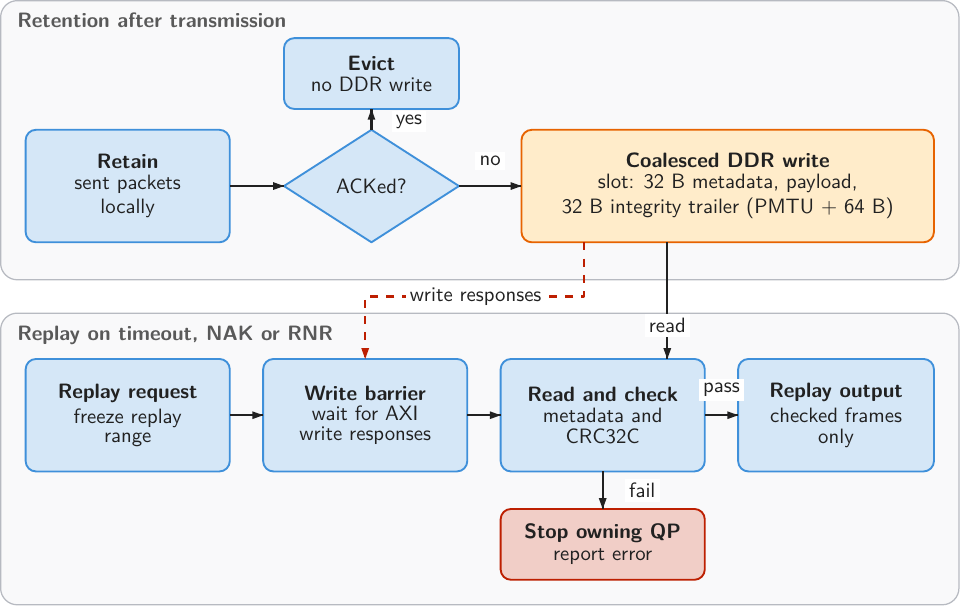}
\caption{Retention and replay logic. Packets acknowledged before writeback are evicted without DDR traffic; a replay waits for the AXI write responses and forwards only frames whose metadata and CRC32C pass.}
\label{fig:3}
\end{figure}

Each retention slot holds 32 B of metadata, one PMTU of payload and a 32 B integrity trailer. At PMTU 1024 B, the 512 MiB DDR region of ZCU102 provides about $4.9\times10^{5}$ slots, divided equally among the active QPs, and the 4 GiB region of the TCB about $3.9\times10^{6}$. New input stalls when the slots are nearly full. A replay request fixes the replay range and waits for the relevant AXI write responses before reading and validating metadata and payload. Only a complete packet that passes these checks enters the replay output (Figure~\ref{fig:3}).

Sequence-error negative acknowledgments (NAKs), RNR responses and timeouts are handled as in the inherited core, with a 5 ms retransmission timeout and three retries without ACK progress. This work adds checks of anomalous ACKs, packet sequence number (PSN) wraparound and the retry budget. An unrecoverable storage error or an exhausted retry budget stops the owning QP and reports error context, after which the host closes or re-establishes the session.

\subsection{Multi-QP operation and flow control}

On ZCU102, records are mapped to up to four QPs by input class, and each QP has its own DDR partition, PSNs and retransmission state. A class change closes the current block, so rotating the class for every record also shortens the blocks.

The extended ZCU102 sender responds to priority-3 PFC pause frames by finishing the frame in progress and blocking the next RoCE frame, and the retransmission timeout is suspended during the pause. With congestion control enabled, outgoing data carry the Explicit Congestion Notification (ECN)-capable codepoint, and the FPGA checks received congestion notification packets (CNPs) and issues per-QP rate events~\cite{RFC3168ECN,Zhu2015DCQCN}. Each QP keeps a congestion estimate and current and target rates, which govern both new data and replay. These extensions are not built into the TCB design.

\FloatBarrier
\section{Evaluation setup and methods}
\label{sec:4}

\subsection{Platforms}

\begin{figure}[htbp]
\centering
\includegraphics[width=\linewidth]{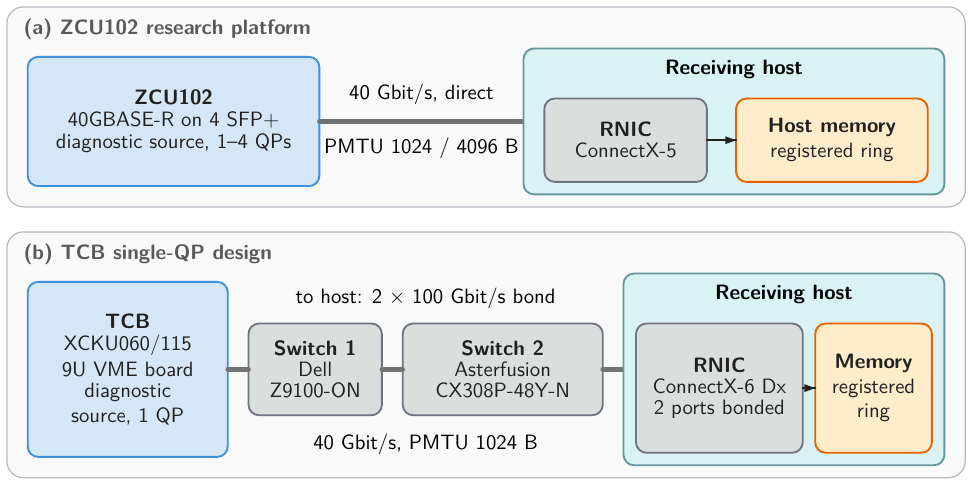}
\caption{Hardware topologies. (a) ZCU102 connected directly to a ConnectX-5 RNIC. (b) TCB connected through two switches to a ConnectX-6 Dx RNIC with two bonded 100 Gbit/s ports. Each board connects to its own receiving host.}
\label{fig:4}
\end{figure}

ZCU102 is connected directly to a ConnectX-5 RNIC. The TCB reaches its receiving host through two switches, a Dell Z9100-ON and an Asterfusion CX308P-48Y-N, and the host uses a ConnectX-6 Dx RNIC whose two 100~Gbit/s ports form one bonded RDMA device (Figure~\ref{fig:4}). Two TCBs are used: an XCKU060 board with the single-QP reference design, and an XCKU115 board with the current TCB firmware. Several ZCU102 builds are used for different studies: aggregation builds at PMTU 1024 and 4096 B, a raw-message build that sends one RDMA message per input frame without aggregation, a pair of retention-policy variants with replay instrumentation, three CRC-configuration variants, and two images with session-management extensions. A further XCKU060 build at PMTU 4096 B serves the SPDK storage prototype. All data come from diagnostic sources.

\subsection{Measurement methods}
\label{sec:4.2}

Payload goodput counts source-record bytes received in complete measurement windows of about 1 s. Where only the receiver's rate log is available (CRC variants, multi-QP rotation, TCB and long transfers), we report this logged transport rate instead. It counts the header and payload bytes of the blocks completed in each window, has 0.01~Gbit/s resolution and exceeds the payload goodput by less than 0.03\% for 1 MiB blocks and by 1.5\% for single-record blocks. Unless stated otherwise, each throughput condition is measured in three sessions of at least 20 GB of source payload. The statistical unit is the independently initialized session. The aggregation builds check the header and payload CRC32C, sequence and notification of every block. The raw-message build relies on rate counters and the RNIC ICRC, with separate captures checked against the diagnostic pattern.

The completion round-trip time (RTT) is measured on the raw-message build. It starts when the FPGA accepts a write request and ends when the retransmission module validates the ACK for the final packet. The next request is issued only after the previous one completes. It is timed in the 312.5 MHz clock domain (3.2 ns per cycle) and includes the transmit pipeline, message transmission, the RNIC response and ACK validation, but not aggregation waiting or host processing. Without the recorder, the raw-message build reports the mean, minimum and maximum of 4096 requests after 128 warm-up requests. For the latency distribution, a passive recorder added to the same build bins every completion: bins are one cycle wide up to 1023 cycles, and each higher power-of-two interval is split into 64 bins. The recorder never stalls the data path and lost no sample. This distribution is measured in five independent sessions of $10^{8}$ requests for each of four conditions (128 B and 4 KiB at PMTU 1024 and 4096 B): quantiles are computed within each session and not pooled.

The same recorder was added to the aggregation build. For every block at full transfer rate it measures the time from the transmission of the last packet to the acknowledgment that covers it, and the time from the arrival of the first record to that acknowledgment, which includes the aggregation wait. Five sessions of 200 GB, about 192,000 blocks each, were run at each of three source rates.

Recovery is tested by suppressing ACKs with a receive gate and by injecting AXI slave-error responses on the DDR read and write paths. For the retention study, two otherwise identical ZCU102 builds differ in one parameter: R-on uses acknowledgment-aware retention, whereas R-off writes every transmitted packet to DDR. A forward injector after the retention logic drops only first-transmission data packets, in bursts of $K$ consecutive packets, at random with probability $p$, or every $N$th packet, and a cycle-timed receive gate delays ACKs. An event recorder timestamps each recovery in clock cycles. For each event, the replay throughput $R_\mathrm{rep}$ is the replayed payload divided by the duration of the replay output, the start latency $L_\mathrm{rep}$ runs from the retry decision to the first replay beat, and the recovery time $T_\mathrm{rec}$ runs from the first dropped packet to the ACK that covers the replayed range. The replay amplification $A$ is the number of replayed packets per dropped packet. The study comprises 86 sessions in 22 conditions. A second matched study of 19 sessions used the same two builds with one record per block.

\FloatBarrier
\section{Performance and validation results}
\label{sec:5}

\subsection{Throughput for detector-sized records}
\label{sec:5.1}

\begin{figure}[htbp]
\centering
\includegraphics[width=\linewidth]{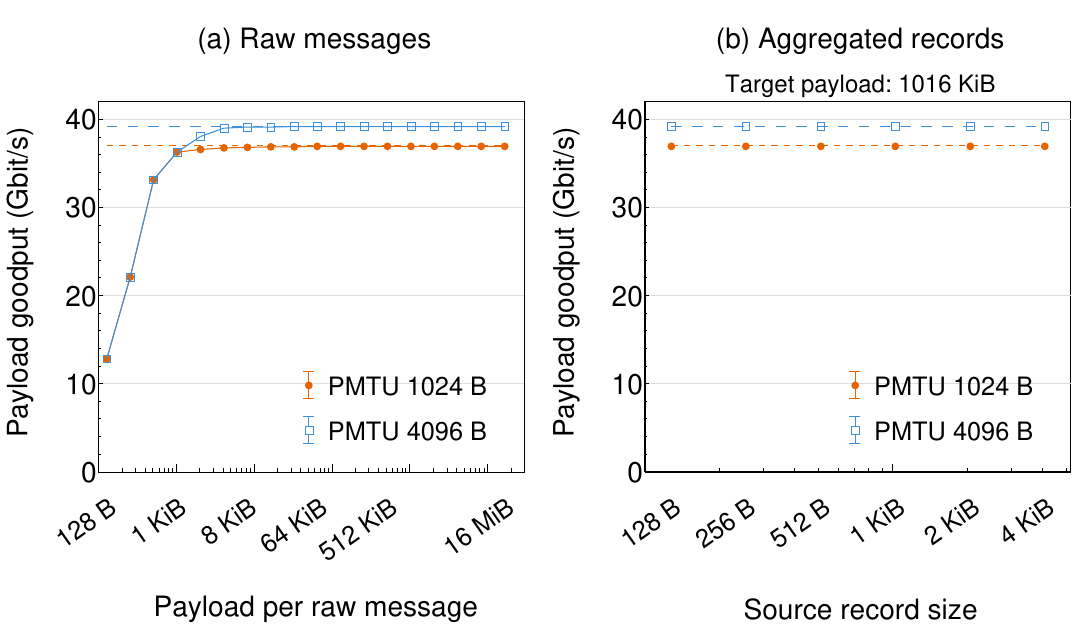}
\caption{Payload goodput for two PMTU settings. (a) Raw messages without aggregation, from 128 B to 16 MiB. (b) Source records from 128 B to 4 KiB aggregated toward a target block payload of 1016 KiB. Each point is the mean of three sessions; bars span the session minimum and maximum. Dashed lines show the packetization limit $R_{\mathrm{env}}$.}
\label{fig:5}
\end{figure}

With source records from 128 B to 4 KiB aggregated toward 1016 KiB blocks, payload goodput is 36.895 and 39.171 Gbit/s at PMTU 1024 and 4096 B, independent of the record length (Figure~\ref{fig:5}(b)). Both values lie within 0.4\% of a simplified packetization limit $R_{\mathrm{env}}=40\,\mathrm{PMTU}/(\mathrm{PMTU}+82)$, which assumes full packets with 82 B of overhead (37.034 and 39.215 Gbit/s). At PMTU 1024 B, the CRC-variant builds (see below) and the TCB (Section~\ref{sec:tcb}) log 37.04~Gbit/s, equal to the limit within the 0.01~Gbit/s resolution, whereas the aggregation build logs 36.90~Gbit/s. The origin of this 0.4\% difference is under study. Because those builds contain the same aggregation and block checks, record handling and checking cost no measurable throughput.

Without aggregation, 128 B messages reach only about 12.8 Gbit/s, a factor of 2.9--3.1 below the aggregated rate, while 16 MiB messages reach 36.901 and 39.177 Gbit/s (Figure~\ref{fig:5}(a)). The two builds also differ in adapters, buffers and checking paths, so these factors compare the two transfer modes as implemented and do not isolate the aggregation logic.

With the class rotated for every record, so that each block holds one record, payload goodput estimated from logged rates is 35.55, 35.69 and 35.69~Gbit/s for 1, 2 and 4 QPs. The per-record rotation itself costs about 3.5\% compared with aggregated blocks. Enabling the CRC of retained records leaves the logged transport rate unchanged: three ZCU102 implementations (disabled, one shared engine, separate write and read engines) all run at 37.04~Gbit/s in normal sessions and recover from ACK gating with two timeouts. The cost of the checks during sustained replay is under study.

\subsection{Latency of the readout path}
\label{sec:5.2}

\begin{figure}[htbp]
\centering
\includegraphics[width=\linewidth]{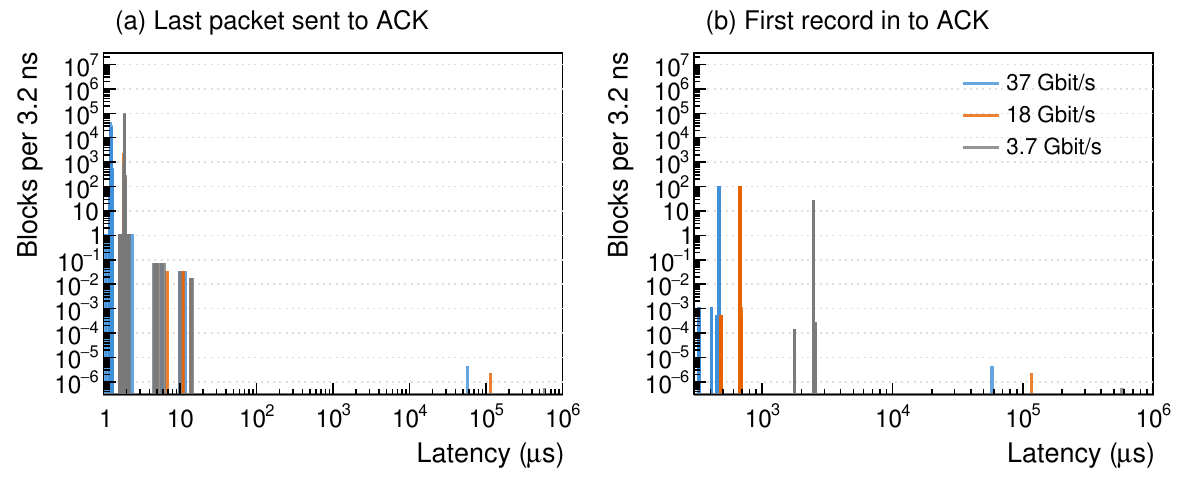}
\caption{Block latency on the aggregation path for 4 KiB records in blocks of about 1 MiB at PMTU 1024 B: (a) from transmission of the last packet of a block to the acknowledgment covering it, and (b) from arrival of the first record of the block to that acknowledgment. Five independent 200 GB sessions per source rate are drawn as separate lines. Bins are 3.2 ns wide up to 3.27~$\mu$s and $1/64$ of an octave above, and counts are normalized to blocks per 3.2 ns.}
\label{fig:l2}
\end{figure}

Figure~\ref{fig:l2} shows the block latency on the aggregation path at source rates of 37, 18 and 3.7~Gbit/s. At full rate the acknowledgment follows the last packet of a block by a median of 1.25~$\mu$s, and in every session 99.99\% of blocks are acknowledged within 1.30~$\mu$s. From the first record of a block to its acknowledgment the latency is 0.47, 0.68 and 2.48~ms at the three rates, with the median and the 99.99th percentile in the same histogram bin. Filling a 1016~KiB block at the source rate and transmitting it at the link rate take 0.225 and 0.226~ms at full rate, together 0.45~ms of the 0.47~ms. On this direct link the acknowledgment adds only microseconds, so an integration can choose the block target from its latency budget, provided a flush rule bounds the wait for sparse input.

\begin{figure}[htbp]
\centering
\includegraphics[width=\linewidth]{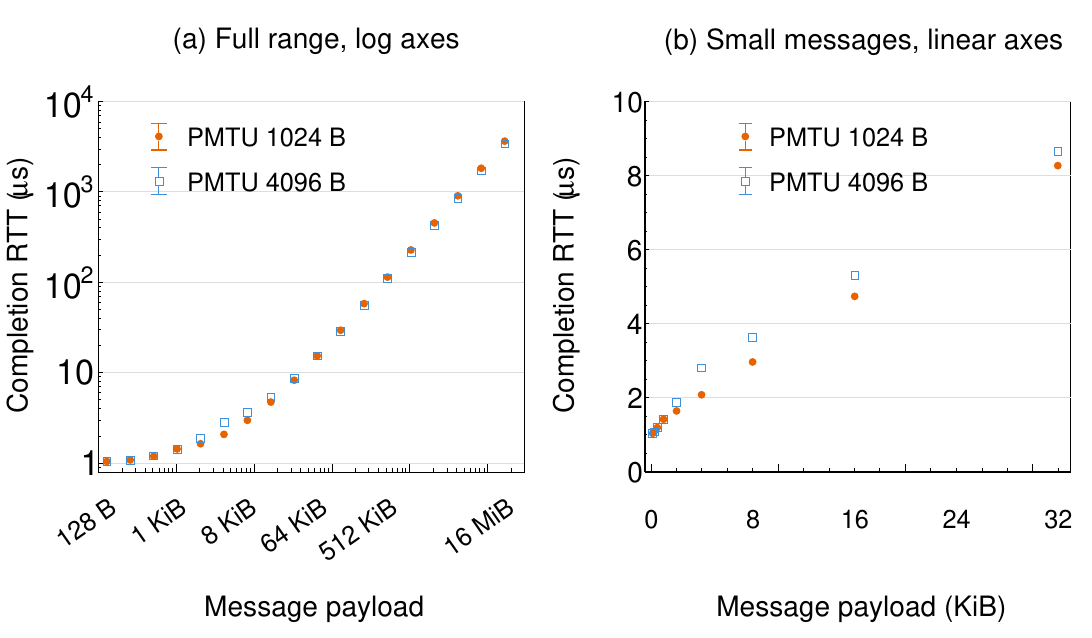}
\caption{Completion RTT of raw messages from FPGA request acceptance to final-ACK validation, excluding aggregation and host processing. Panel (a) covers 128 B to 16 MiB; (b) shows 128 B to 32 KiB on linear axes. Each condition is one session; points show means of 4096 serial requests and bars their minimum and maximum, which are mostly smaller than the markers.}
\label{fig:6}
\end{figure}

The microsecond part is the completion round trip of the transport itself, which the raw-message path measures without aggregation (Figure~\ref{fig:6}). Its mean is 1.034~$\mu$s for 128 B, and 2.077 and 2.810~$\mu$s for 4 KiB at PMTU 1024 and 4096 B. These values mainly characterize the inherited protocol core with the 40 Gbit/s port.

\begin{figure}[htbp]
\centering
\includegraphics[width=\linewidth]{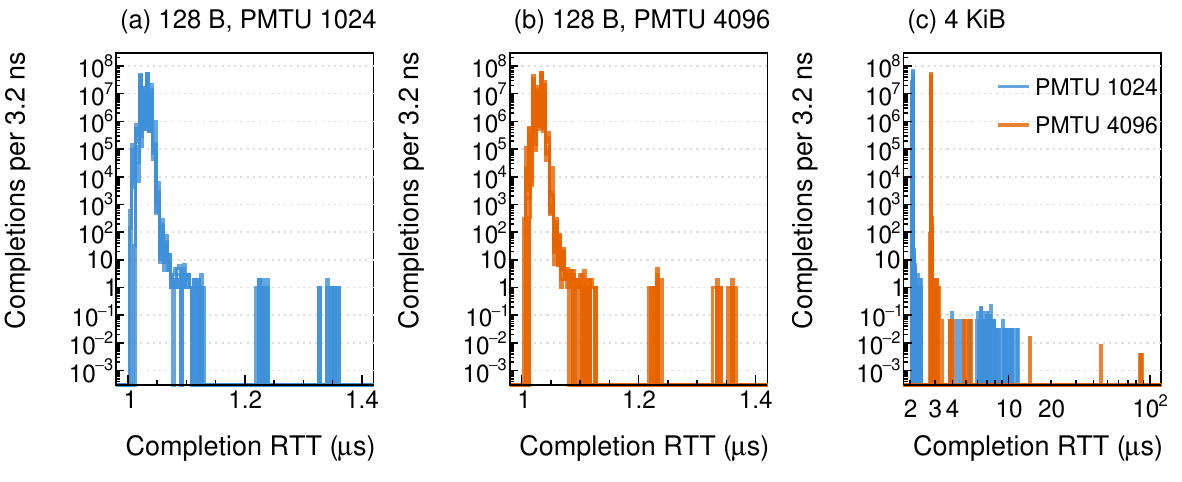}
\caption{Completion-RTT histograms of 128 B requests at PMTU (a) 1024 B and (b) 4096 B, and (c) of 4 KiB requests at both PMTUs. Each condition has five independent sessions of $10^{8}$ serial requests, drawn as separate lines. Bins are 3.2 ns wide up to 3.27~$\mu$s and $1/64$ of an octave above; counts are normalized to completions per 3.2 ns, so that bins of different width are comparable.}
\label{fig:taillat}
\end{figure}

The recorder added to the same path gives the full distribution (Figure~\ref{fig:taillat}). Over $10^{8}$ serial requests per session, 128 B completions have a median of 1.02--1.03~$\mu$s, a 99.999th percentile of at most 1.053~$\mu$s at both PMTUs and a maximum of 1.35--1.37~$\mu$s. For 4 KiB requests the median is 2.09 and 2.81~$\mu$s at PMTU 1024 and 4096 B, and the 99.999th percentile is 19 and 45~ns higher. Every 4 KiB session, however, also contains rare long completions of up to 88~$\mu$s, which the mean, minimum and maximum of 4096 requests do not reveal. Above fixed thresholds of 4.2 and 5.7~$\mu$s, there are 135 such completions at PMTU 1024 B and 5 at PMTU 4096 B in $5\times10^{8}$ requests per condition. Their origin is under study.

\subsection{Recovery and fault containment}
\label{sec:5.3}

\begin{figure}[htbp]
\centering
\includegraphics[width=\linewidth]{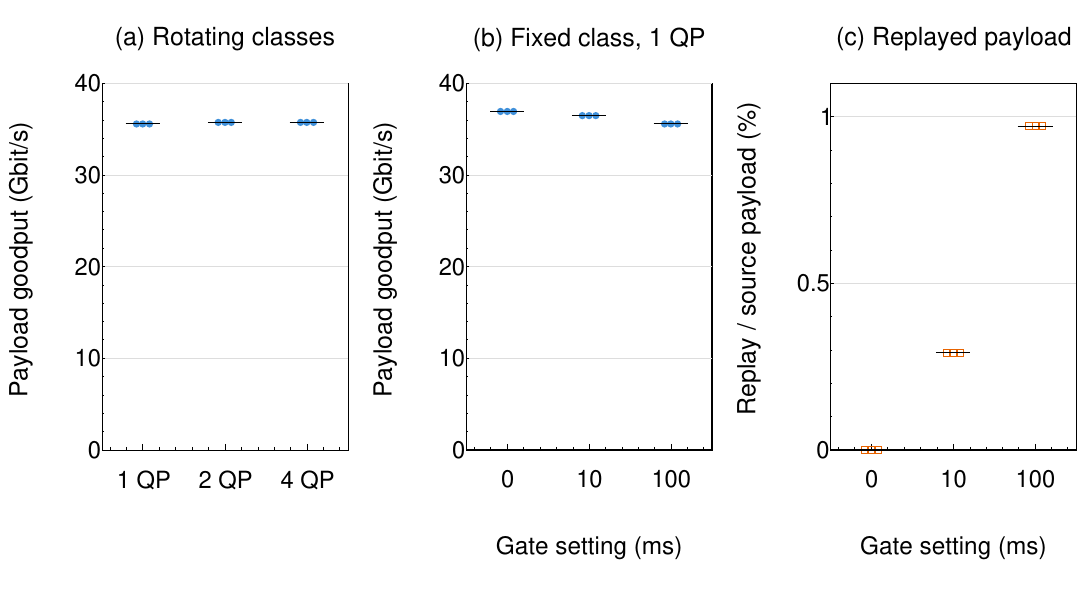}
\caption{Multi-QP and controlled-replay results at PMTU 1024 B. (a) Payload goodput, estimated from logged window rates, with rotating-class input for each QP count. (b) Payload goodput for single-QP traffic at gate settings of 0, 10 and 100 ms. (c) Replayed payload as a percentage of source payload for the same sessions as (b). Markers show three sessions per condition and horizontal lines their means.}
\label{fig:7}
\end{figure}

Recovery and fault containment were tested with the aggregation build at PMTU 1024 B and 4 KiB records. With ACKs suppressed by script-controlled gate settings of 10 and 100 ms, the timeout replays the outstanding data, and every session completes with all host checks passing. Payload goodput falls from 36.898 to 36.485 and 35.540~Gbit/s (Figure~\ref{fig:7}(b,c)).

DDR access errors stayed within their QP. With four QPs active on ZCU102, which run at the same goodput as one (Figure~\ref{fig:7}(a)), a write-response error on one QP stops and reports that QP, and a read-response error blocks the unchecked replay and takes only the affected QP out of service. In both cases the other QPs complete their transfers with all checks passing. Each fault type was injected in eight independent sessions, five on one QP and three on another, and all 16 injections stayed within their QP. Corruption of retained metadata or payload, anomalous ACKs, PSN wraparound and retry exhaustion were tested in directed RTL simulation.

\subsection{Cost of retention and behavior of replay}
\label{sec:c8z-results}

All 86 sessions of the matched R-on/R-off study passed their checks, with converged PSNs and no retry exhaustion.

The R-off variant shows the cost that the policy avoids. Without injected loss, R-on writes 0.028\% of the source payload to the DDR interface and delivers 36.90~Gbit/s; R-off writes 106.3\% and delivers 31.03~Gbit/s (five sessions each). The R-off write interface was stalled in about 45\% of its valid cycles, so its lower goodput reflects the DDR write bandwidth of this board. The traffic saving depends instead on acknowledgments returning before retained data must be written back: a 1~ms receive gate, which delays ACKs without causing replay because it is shorter than the 5~ms timeout, already raises the R-on fraction to 0.135\%, and longer delays in switched networks are expected to raise it further. Under random loss the goodput difference persists at $p=10^{-4}$ (36.82 versus 30.99~Gbit/s) and nearly vanishes at $p=10^{-3}$ (28.44 versus 28.24~Gbit/s), where recovery stalls dominate.

With one record per block (second study, 19 sessions; five packets per 4~KiB block), R-on wrote 33--42\% of the payload to DDR (median 40\%, five sessions) and R-off 109\% (three sessions), with goodput of 35.54 and 28.93~Gbit/s. A 100~$\mu$s receive gate gave 35--42\%. The same study reproduced the large-block values (0.028--0.030\% and 106\%). The mechanism behind this fraction will be studied in future work.

\begin{figure}[htbp]
\centering
\includegraphics[width=\linewidth]{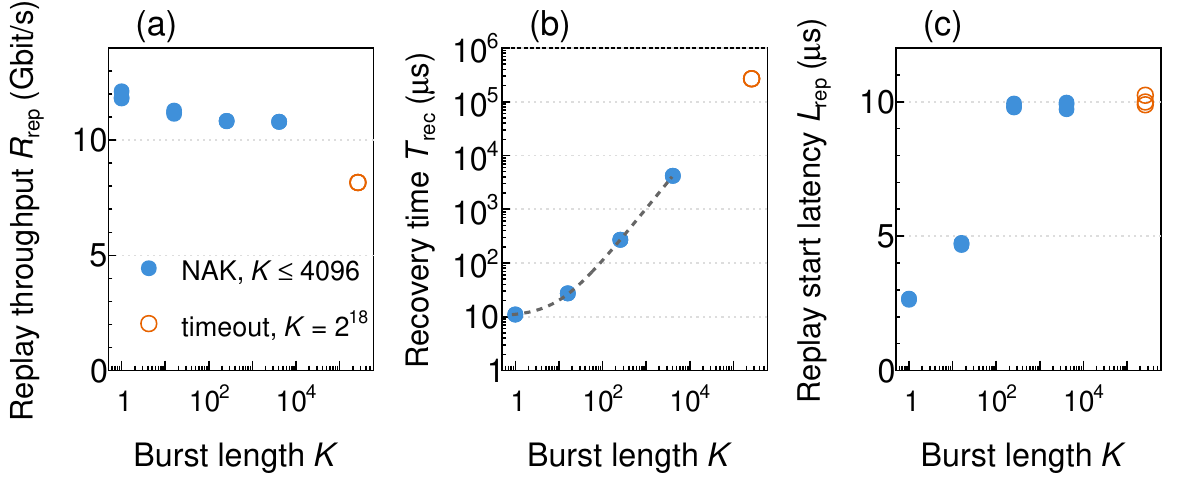}
\caption{Replay path for bursts of $K$ dropped first-transmission packets (R-on, four bursts per session, three sessions per $K$). (a) Replay throughput $R_\mathrm{rep}$, (b) recovery time $T_\mathrm{rec}$ with the least-squares line for $K\le4096$ (1.02~$\mu$s per packet), (c) replay start latency $L_\mathrm{rep}$. Each marker is a session median. Open markers ($K=2^{18}$) are timeout-driven recovery and are not part of the fit.}
\label{fig:c8z-replay}
\end{figure}

For bursts of up to $K=4096$ dropped packets, a sequence-error NAK triggers recovery, replay runs at 10.8--11.9~Gbit/s, and $T_\mathrm{rec}$ grows by 1.02~$\mu$s per packet of the burst with an offset of 10~$\mu$s (Figure~\ref{fig:c8z-replay}). The slope combines the time the burst occupies the link (about 0.22~$\mu$s per 1~KiB packet) with the replay of each packet (about 0.8~$\mu$s). The start latency rises from 2.7~$\mu$s at $K=1$ to about 9.9~$\mu$s for large bursts. Bursts longer than the number of packets sent in one 5~ms timeout interval (about $2\times10^{4}$), such as $K=262144$, are recovered by timeouts, and goodput falls to 30.7~Gbit/s. The factors that limit replay to about 28\% of line rate are under study.

\begin{figure}[htbp]
\centering
\includegraphics[width=\linewidth]{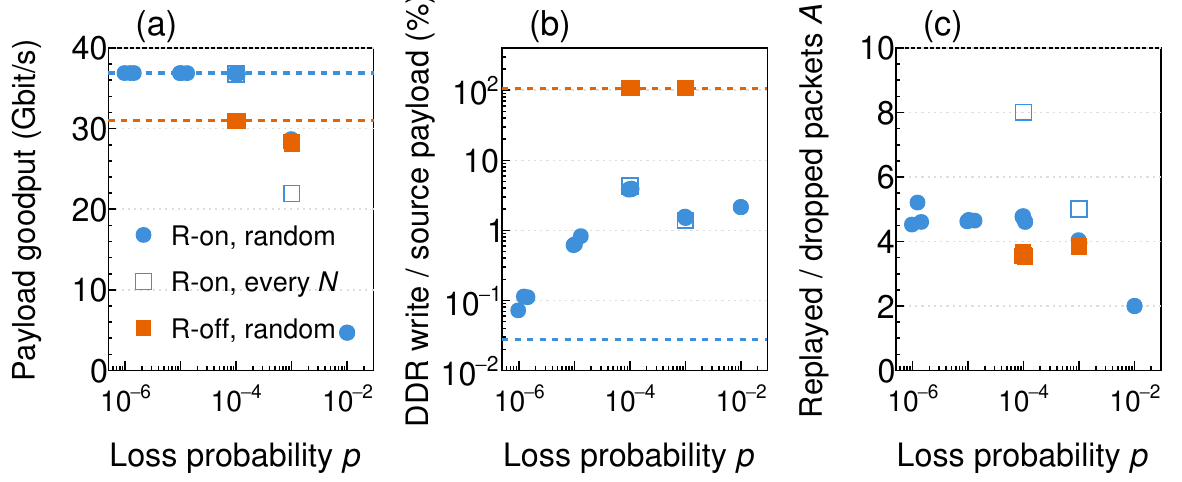}
\caption{Sessions with injected loss against the observed loss probability. (a) Payload goodput, (b) DDR write-interface bytes as a percentage of source payload, (c) replay amplification $A$. Open squares drop every $N$th packet ($N=10^4$, $10^3$). Dashed lines are the no-injection means of R-on (blue) and R-off (orange). Each marker is one session.}
\label{fig:c8z-loss}
\end{figure}

Under random loss, goodput stays within about 0.2\% of the loss-free value up to $p=10^{-4}$ and falls to 28.4~Gbit/s at $10^{-3}$ and 4.70~Gbit/s at $10^{-2}$ (Figure~\ref{fig:c8z-loss}). The condition means of the replay amplification $A$ are 3.6--4.8 up to $p=10^{-3}$, as expected for go-back-$N$ recovery that resends the unacknowledged packets after each loss: a completion round trip of about 1~$\mu$s spans about 4.7 packets of 0.22~$\mu$s. The replayed volume is only about 0.4\% of the payload at $p=10^{-3}$, so the goodput loss comes from recovery stalls. Per dropped packet, the loss corresponds to about 5~$\mu$s of transmit time up to $p=10^{-4}$ and 66~$\mu$s at $10^{-3}$.

The loss rate observed on the trigger board, about $3\times10^{-10}$ per packet (Section~\ref{sec:tcb}), is more than five orders of magnitude below $10^{-4}$, the highest injected probability at which goodput stayed within about 0.2\%. The comparison assumes independent losses; losses in a switched network may be correlated.

\subsection{Operation on the trigger board}
\label{sec:tcb}

Three TCB builds were tested, all through the two-switch network of Figure~\ref{fig:4}(b): the single-QP reference design on an XCKU060 board, the current TCB firmware on an XCKU115 board, and a PMTU 4096 B XCKU060 build for the storage prototype (Section~\ref{sec:storage}). On the XCKU060 board the reference design transferred 20 GB in a normal session and in a session with ACK gating. In the latter, the data were recovered by two timeout replays and all checks passed.

All 45 sessions on the XCKU115 board passed (Table~\ref{tab:c10}). As on ZCU102, the rate does not depend on record length. The logged transport rate is 37.04~Gbit/s from 64~B to 4~KiB. Reduced source rates of 27.82, 18.55, 9.28 and 3.70~Gbit/s, about 75, 50, 25 and 10\% of full rate, were carried without loss. In three transfers of 16.4~TB, each lasting about one hour and carrying about $1.6\times10^{10}$ packets, 4, 5 and 5 packets were lost between the FPGA MAC and the RNIC, about $3\times10^{-10}$ per packet. Every loss was recovered by NAK-triggered replay without timeout.

\begin{table}[htbp]
\small
\caption{TCB with XCKU115 (PMTU 1024~B, single QP). Rates are logged transport rates including block headers, at 0.01~Gbit/s resolution; $n$ is the number of independent sessions.}
\label{tab:c10}
\begin{tabularx}{\linewidth}{@{}YrYY@{}}
\toprule
Test & $n$ & Result & Loss and recovery \\
\midrule
Record length 64--4096 B, 7 values & 21 & 37.04 Gbit/s at every length & none \\
Inter-record gap, 5 settings & 15 & 37.04, 27.82, 18.55, 9.28, 3.70 Gbit/s & none \\
Long transfer, 16.4 TB each & 3 & 37.04 Gbit/s & 4, 5, 5 packets; replayed, no timeout \\
CLOSE and reopen, 3 pairs & 6 & content checks pass & none \\
\bottomrule
\end{tabularx}
\end{table}

\subsection{Long transfers, storage and session lifecycle}
\label{sec:storage}

The two ZCU102 images with session-management extensions were used for long transfers and session tests. Each completed a 500 GB transfer with 4 KiB records rotating among four QPs at PMTU 1024 B, at a logged transport rate of 36.25~Gbit/s and without NAK, timeout or replay. The final counts imply, per QP, nine wraparounds of the 24-bit PSN and one of the 16-bit generation that the Immediate carries to distinguish successive passes through the host ring. Both passed without error. In seven active-CLOSE cases per image, with one, two or four QPs, the session was closed while the source was still busy, all QPs reached RESET, and a new session opened without reprogramming completed a 20 GB transfer with full count closure. Data in flight at the forced close are not guaranteed to arrive.

For storage, the SPDK prototype on the XCKU060 board wrote 40 GB of source payload to an NVMe device and passed full readback with block CRC and sequence checks in a single session. Storage backpressure reached the sender as 693 RNR events and 7,508 replayed packets and lowered the rate from 39.22~Gbit/s in the first windows to 35.95--36.81~Gbit/s, and the transfer completed.

\subsection{Flow and congestion control}
\label{sec:flow-control-results}

PFC reception was tested in hardware and the DCQCN rate controller in simulation. Six hardware cases on a direct link checked PFC: in the two priority-3 cases transmission paused and resumed after release or expiry and the transfer completed, and four negative controls produced no pause. The DCQCN controller was studied with directed RTL tests and a packet-level model that compiles the actual CNP parser and rate-controller RTL and represents eight 40~Gbit/s sources sharing one or three 100~Gbit/s links. All 125 model runs passed their checks. For mixed packet lengths the mean Jain fairness index is $0.74\pm0.04$ against a target of 0.95. Tuning of the rate-control profile and closed-loop tests with ECN marking by a physical switch are under study.

\FloatBarrier
\section{Target application: BESIII MDC}
\label{sec:6}

The MDC application aims to acquire the complete trigger input within the connected coverage for offline replay and algorithm studies while preserving the existing trigger functions~\cite{Wang2026MDCReadout}. Each TCB receives eight 10 Gbit/s serial links from the MDC back-end electronics and connects to a DAQ server through two switches by 40 Gbit/s RoCEv2, and the DAQ server forwards the data to a storage server through XRootD~\cite{IHEPTDAQDRD72026}. The TCB hardware provides two 40 Gbit/s ports, of which the application uses one, as in the design evaluated here. One channel has operated in beam and cosmic-ray tests, and eight installed TCBs, each reaching 13.4 Gbit/s without zero suppression, are under test~\cite{Dong2026CEPCRoCE,IHEPTDAQDRD72026}. For MDC data, the integration defines the record lengths and delimiters and the receiving software the buffer handling. The transport itself is used unchanged.

\FloatBarrier
\section{Conclusions}
\label{sec:7}

A 40 Gbit/s FPGA RoCEv2 transport has been developed to export BESIII MDC trigger inputs to server memory. It runs on a ZCU102 board and as a single-QP design on TCBs with XCKU060 and XCKU115 devices. All measurements use diagnostic data sources.

Throughput is set by the link. With records aggregated into 1 MiB blocks, payload goodput comes within 0.4\% of the packetization limit, and builds with the full block checks reach the limit within the log resolution. Recovery replays lost or unacknowledged data, and goodput falls by only about 0.2\% at injected loss rates far above those observed on the trigger board; DDR access errors stay within their queue pair.

The cost of retention depends on how quickly acknowledgments return. On a direct link, acknowledgment-aware retention removes almost all DDR writes for large blocks, while single-record blocks still write about 40\% of their payload. Its benefit in switched networks will be measured with the installed system. Block latency is set by block size and input rate, so an integration can choose the block size from its latency budget.

On the trigger board, the transport carried three 16.4~TB transfers of about one hour each through a switched network, and every one of the rare packet losses was recovered by replay. An installation at the BESIII MDC is under test~\cite{IHEPTDAQDRD72026}, and validation with recorded detector data is the next step.

\section*{Acknowledgments}

This work was supported by the Young Scientists Fund of the National Natural Science Foundation of China under Grant No.~12505223.

\FloatBarrier
\section*{Declaration of competing interest}

The authors declare that they have no known competing financial interests or personal relationships that could have appeared to influence the work reported in this paper.

\section*{Data availability}

Data will be made available on request.

\section*{Declaration of generative AI and AI-assisted technologies in the writing process}

During the preparation of this work the authors used Claude (Anthropic) to translate the authors' Chinese manuscript into English and to edit and rephrase the English text in formal writing language. After using this tool, the authors reviewed and edited the content as needed and take full responsibility for the content of the publication.

\FloatBarrier
\bibliographystyle{elsarticle-num}
\bibliography{references_elsarticle}

\end{document}